\documentclass[aps,pra,amsmath,twocolumn,amssymb,footinbib,superscriptaddress]{revtex4-1}

\usepackage{amssymb,amsfonts,amsmath}

\usepackage{color}
\usepackage[apple mac]{inputenc}
\usepackage[english]{babel}
\usepackage{times}
\usepackage{latexsym}
\usepackage{fancyhdr}
\usepackage{verbatim}
\usepackage{graphicx}
\usepackage{epsf}
\usepackage{subfigure}
\usepackage{bm}
\usepackage{epstopdf}\definecolor{Nathanblue}{rgb}{0.,0.24,0.51}

\def\be{\begin{equation}}
\def\ee{\end{equation}}
\def\bs#1{\boldsymbol{#1}}
\def\txt#1{\textrm{#1}}

\newcommand{\eps}{\varepsilon}
\renewcommand{\i}{\mathrm{i}}
\newcommand{\e}{\mathrm{e}}
\newcommand{\ba}{\bm{a}}
\newcommand{\bb}{\bm{b}}
\newcommand{\bk}{\bm{k}}
\newcommand{\bp}{\bm{p}}
\newcommand{\br}{\bm{r}}

\newcommand{\lam}{\lambda_{\text{am}}}

\begin{document}

\title{Design of laser-coupled honeycomb optical lattices supporting Chern insulators}

\author{E. Anisimovas}
\email[]{egidijus.anisimovas@ff.vu.lt}
\affiliation{Department of Theoretical Physics, Vilnius University, Saul\.{e}tekio 9, LT-10222 Vilnius, Lithuania}
\affiliation{Institute of Theoretical Physics and Astronomy, Vilnius University, Go\v{s}tauto 12, LT-01108 Vilnius, Lithuania}

\author{F. Gerbier}
\affiliation{Laboratoire Kastler Brossel, CNRS, UPMC, ENS, 24 rue Lhomond, 75005, Paris, France}

\author{T. Andrijauskas}
\affiliation{Institute of Theoretical Physics and Astronomy, Vilnius University, Go\v{s}tauto 12, LT-01108 Vilnius, Lithuania}

\author{N. Goldman}
\email[]{nathan.goldman@lkb.ens.fr}
\affiliation{Laboratoire Kastler Brossel, CNRS, UPMC, ENS, 24 rue Lhomond, 75005, Paris, France}

\date{\today}

\begin{abstract}
We introduce an explicit scheme to realize Chern insulating phases employing cold atoms trapped in a state-dependent optical lattice and laser-induced tunneling processes. The scheme uses two internal states, a ground state and a long-lived excited state, respectively trapped in separate triangular and honeycomb optical lattices. A resonant laser coherently coupling the two internal states enables hopping between the two sublattices with a Peierls-like phase factor. Although laser-induced hopping by itself does not lead to topological bands with non-zero Chern numbers, we find that such bands emerge when adding an auxiliary lattice that perturbs the lattice structure, effectively turning it at low energies into a realization of the Haldane model: A two-dimensional honeycomb lattice breaking time-reversal symmetry. We investigate the parameters of the resulting tight-binding model using first-principles band structure calculations to estimate the relevant regime for experimental implementation.
\end{abstract}

\maketitle

\section{Introduction}

Lattice systems displaying topologically nontrivial band structures are currently attracting the curiosity of a large scientific community~\cite{Hasan2010,Qi2011}.  For systems breaking time-reversal invariance, the band topology is characterized by a topological invariant (the Chern number~\cite{Thouless1982,Kohmoto:1985}) taking integer values. The presence of topological order is signaled by a non-zero value of the topological invariant and has experimental consequences, such as the existence of chiral edge states enforced by the bulk-edge correspondence~\cite{Hasan2010,Qi2011} or the quantization of transport coefficients in electronic systems~\cite{vonKlitzing:1986}. The universality of these topological properties suggests that they could be engineered not only in solid-state systems~\cite{Qi2011} but also in a wide range of physical systems characterized by spatially periodic structures, such as photonic lattices~\cite{photonic} or ultracold atoms trapped in optical lattices~\cite{Bloch:2008,Dalibard2011,Goldman:2013review}. Progress towards realization of topological phases in cold atomic gases has been recently reported~\cite{Aidelsburger:2011,Aidelsburger:2013,Ketterle:2013}. 

\begin{figure*}
\centering
\includegraphics[width=\textwidth]{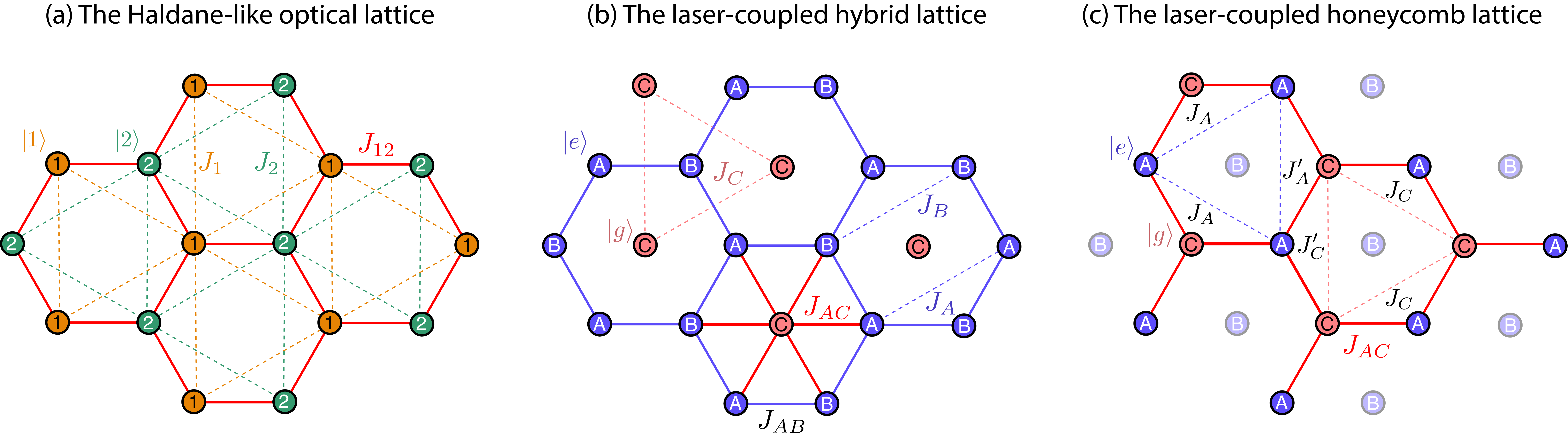}
\caption{\label{fig-intro} (Color online) (a) The Haldane-like optical lattice: Atoms in states $\vert 1 \rangle$ and $\vert 2 \rangle$ are trapped in two state-dependent triangular optical lattices, with hopping amplitude $J_1$ and $J_2$, respectively. Coupling the states $1 \leftrightarrow 2$ induces hopping between the two triangular lattices, generating a Haldane-like honeycomb lattice with complex NN hopping matrix element $J_{12}$. (b) Sketch of the hybrid laser-coupled lattice generated by the potentials in Eq.~\eqref{eq-lg1}. Atoms in the ground $g$ (resp.\ excited $e$) state are attracted to the intensity maxima (resp.\ minima) that span the red triangular (resp.\ blue honeycomb) sublattice. Full and dotted  lines indicate the dominant NN and NNN tunnelings, which enter the tight-binding models of Section~\ref{section:main_lattices}. The two sublattices are coupled by direct laser-induced transitions indicated with full red lines, as described in Eq.~\eqref{hopping_phase}. (c) Sketch of the final lattice geometry resulting from the addition of an auxiliary lattice (Section~\ref{auxiliary}), where B sites are blacked out to indicate that they correspond to higher bands of the hybrid lattice. A and C sites form a honeycomb lattice connected by NN laser-induced tunneling (full red lines) and natural NNN tunneling (dashed lines). Note that the NNN hopping amplitudes are slightly anisotropic: $J_{A} \ne J_A'$ and $J_{C} \ne J_C'$. The laser-coupled honeycomb lattice in (c) is qualitatively equivalent to the Haldane-like optical lattice (a) of Refs.~\cite{Alba:2011,Goldman:2013njp}.}
\end{figure*}

In two space dimensions, one of the simplest models supporting topological bands was proposed by Haldane~\cite{Haldane:1988}. This model features nearest-neighbour (NN) and \emph{next}-nearest-neighbour (NNN) hoppings on a honeycomb lattice accompanied with complex (Peierls) phase factors such that the net flux through a unit hexagonal cell is zero. The Haldane model has never been realized in laboratories, but it has been suggested that it could be engineered through lattice shaking~\cite{Hauke:2012}, rotation \cite{Wu:2008} or laser-induced methods~\cite{Shao:2008,Hauke:2012,Liu:2010,Stanescu2010,Goldman:2013Haldane}. In the present contribution, we consider a concrete experimental implementation of the Haldane-like optical lattice, initially introduced by Alba et al. \cite{Alba:2011}. This scheme, illustrated in Fig.~\ref{fig-intro}(a), envisages trapping atoms with two internal states into two state-dependent triangular optical lattices. The two lattices are spatially distinct and intertwined to form a honeycomb pattern. A laser-induced  coupling of the two internal states produces the NN hoppings within this ``hybrid honeycomb'' lattice, as shown by full red lines in Fig.~\ref{fig-intro}(a). The properties of this model were explored in detail in Ref.~\cite{Goldman:2013njp} in terms of a simple two-band tight-binding model. However, the relation between the tight-binding model parameters and the realistic lattice potential was not explored in the previous works~\cite{Alba:2011,Goldman:2013njp}.

In this Article, we go beyond the studies \cite{Alba:2011,Goldman:2013njp} and analyze an explicit experimental scheme that implements the Haldane-like optical lattice. We build on the scheme proposed in Ref.~\cite{Gerbier:2010}, and consider an atomic species with a long-lived metastable excited state $e$. The method is valid both for bosonic and fermionic species, and it does not suffer from spontaneous emission. We show that the Haldane-like optical lattice  can be realized using a minimal set of ingredients: (a) a primary state-dependent lattice $V^{(1)}$, that traps the ground $g$ and excited $e$ states in a honeycomb/triangular hybrid lattice; (b) a laser that couples the two states $g \leftrightarrow e$, and (c) an auxiliary lattice $V^{(2)}$ periodic in one direction only. The main lattice $V_{g/e}^{(1)}$, depicted in Fig.~\ref{fig-intro}(b), traps the ground state $g$ in the triangular lattice represented by the C sites, while the excited state $e$ is trapped in the complementary honeycomb lattice formed by the A and B sites. Additionally, the $g$ and $e$ states are coupled by a resonant laser inducing hopping between the primary honeycomb and triangular lattices. Superimposing an auxiliary lattice $V_{g/e}^{(2)}$ chosen to shift the B sites in energy, one effectively removes these sites from the dynamics. The resulting ``laser-coupled honeycomb'' lattice is illustrated in Fig.~\ref{fig-intro}(c). It is characterized by laser-induced NN hopping between A and C sites, with complex tunneling matrix elements $J_{AC} \,\e^{\i \bs p \cdot (\bs r_{A}+ \bs r_C)/2}$, and natural NNN hopping between sites of same nature, with amplitudes $J_{A} \approx J_C$. This laser-coupled honeycomb lattice is qualitatively equivalent to the Haldane-like model of Refs. \cite{Alba:2011,Goldman:2013njp} which was shown to host Chern insulating phases for specific values of the transferred momentum $\bs p$ and ratio $J_{A}/J_{AC}$. Thus, the present scheme illustrated in Fig.~\ref{fig-intro}(c) provides a realistic method to realize topological bands in optical-lattice systems. In the following, we investigate this strategy in detail and discuss its validity in terms of actual lattice parameters.

The present work is structured as follows. In Section \ref{section:main_lattices}, we discuss the atomic properties of two-electron atoms used in our proposal, taking the specific example of Ytterbium atoms, introduce the main lattice $V^{(1)}$, and explore its band structure. In Subsection \ref{section:coupling}, we discuss the lattice structure emerging from the coupling between the ground $g$ and excited $e$ states. Then we add the auxiliary lattice $V^{(2)}$ in Section \ref{auxiliary}, and we show how it leads to a Haldane-like model exhibiting reachable Chern insulating phases. We conclude with a summary and some final remarks in Section \ref{conclusion}.

\section{Hybrid triangular-honeycomb lattice}
\label{section:main_lattices}

\subsection{Atomic structure and light-shift potentials}

We consider a gas of atoms with two internal states, denoted $g $ and $e$, which are trapped in a potential landscape created by a set of lasers. A key requirement is to choose a long-lived excited state $e$ to suppress heating due to spontaneous emission. This is for instance fulfilled in alkaline-earth or Ytterbium atoms~\cite{Gerbier:2010} where $g$ is chosen to be the electronic spin singlet ground state $^1 S_0$, and $e$ is a long-lived spin triplet excited state $^3 P_0$. The transition between these two states has already been exploited to build atomic clocks \cite{barber2006a,lemke2009a}, and proposed to be a good candidate for coherent operations in quantum information processing \cite{yi2008a} or quantum simulation \cite{Gerbier:2010,gorshkov2010a}. In the following we choose Ytterbium atoms to be specific while the proposed method should work as well with other atoms featuring very long-lived excited states. For Ytterbium, the lifetime of the $^3 P_0$ excited state is estimated to be $\sim 20\,\text{s}$~\cite{porsev2004a}, and coupling to the ground state is achieved using a laser at the resonant wavelength $\lambda_{ge} \approx 578\,\text{nm}$.  

We consider here atoms confined to two dimensions by a strong trap in the $z$ direction acting identically on both internal states. In general, the potentials $V_{g/e}(\bs r)$ felt by the two states are different~\cite{Grimm:2000}. For the sake of simplicity, we choose the so-called ``anti-magic'' wavelength $\lam$ at which the polarizabilities of the two relevant states are exactly opposite, $\alpha_g (\lam)= - \alpha_e (\lam)=\alpha_{\text{ am}}>0$~\cite{yi2008a,Gerbier:2010}. Generalizing to another wavelength is straightforward as long as the signs of the polarizabilities remain opposite. For a monochromatic laser, the optical lattice potential $V_{g/e}^{(1)} (\bs r)$ felt by atoms in each state $g/e$ can then be written in terms of the total electric field ${\bs E} (\br)$ as~\cite{Grimm:2000}
\be 
\label{eq:lightshift}
V_{g/e} (\bs r)= \mp \frac{1}{2} \alpha_{\text{am}} \vert {\bs E} (\br) \vert^2.
\ee 
Ground state atoms are trapped near the maxima of the intensity $\propto \vert {\bs E} (\br) \vert^2$, while excited state atoms are trapped near the minima. Importantly, the anti-magic wavelength should be far detuned from any resonance so as to avoid spontaneous emission in the experiment. In the following, we will consider optical lattices at the Yb anti-magic wavelength $\lam\approx 1120\,\text{nm}$~\cite{Gerbier:2010,dzuba2010a}. The energy will thus be measured in units of the recoil energy $E_R/h=\left ( h/2m \lam^2 \right ) \approx 900\,\text{Hz}$, corresponding to a temperature of about $T_R \approx 40\,\text{nK}$. 

\begin{figure*}
\centering
\includegraphics[width=\textwidth]{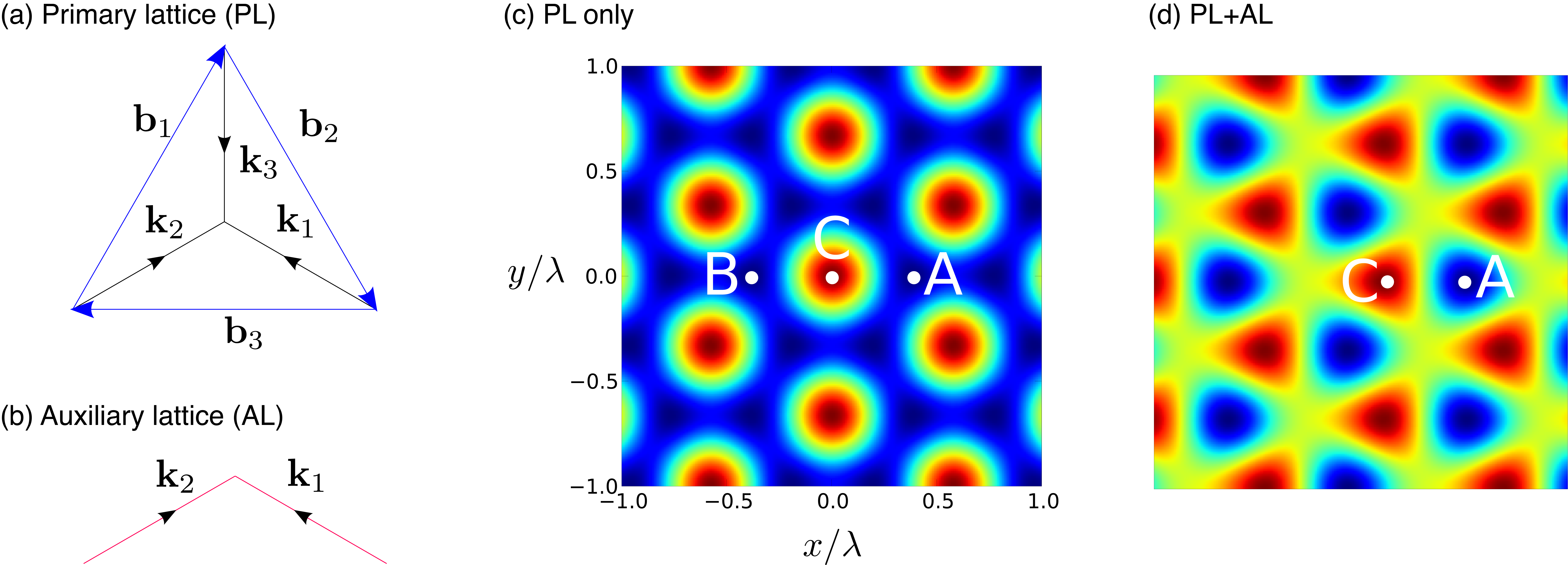}
\caption{\label{fig-lattice} (Color online) (a) Laser beams configuration for the primary lattice:
The wavevectors $\bs k_{1,2,3}$ of the three lasers and the elementary reciprocal lattice vectors $\bb_{1,2,3}$ are shown. (b) Laser beams configuration for the auxiliary lattice. Note that primary and auxiliary fields add incoherently due to the choice of polarizations. (c) The intensity distribution $\vert \bs E (\bs r) \vert^2$ for the primary lattice: Red (resp. blue) colors correspond to high (resp. low) intensity regions. The inequivalent lattice sites of the corresponding honeycomb (A, B) and triangular (C) lattices are indicated. (d) The intensity distribution $\vert \bs E (\bs r) \vert^2$ for the total lattice potential obtained by adding the auxiliary lattice. The inequivalent lattice sites of the corresponding honeycomb  lattice (A, C) are indicated.}
\end{figure*}

\subsection{Band structure calculation for the primary lattices}

The primary lattice is created by three coplanar laser beams of equal wavelength $\lam$ and intensity, and wave vectors ${\bs k}_{1,2,3}$ that intersect at angles $2\pi /3$ [see Fig.~\ref{fig-lattice}(a)]. We choose the polarizations of the beams to be normal to the lattice plane, so that the complex amplitudes of the corresponding electric fields take the form ${\bs E}_{i} = E_0 {\bf e}_z \, \e^{\i\bk_i\cdot\br}$. The coherent superposition of the individual electric fields results in the intensity distribution illustrated in Fig.~\ref{fig-lattice}(c). The intensity maxima (resp.\ minima) of this pattern lie on a triangular (resp.\ honeycomb) lattice that traps $g$ (resp.\ $e$) atoms according to Eq.~(\ref{eq:lightshift}) \cite{grynberg2001a,becker2010a}, as illustrated in Fig.~\ref{fig-intro}(b). In the following we label A and B the two inequivalent sites of the honeycomb lattice formed by intensity minima, and C the sites of the triangular lattice formed by intensity maxima. We write the light-shift potentials acting on $g/e$ as
\begin{equation}
\label{eq-lg1}
V_{g/e} ^{(1)}({\bs r})= \mp V_0 \left[3 + 2 \sum_{j=1}^3 \cos \left(\bb_j \cdot \br\right)\right],
\end{equation}
where we introduced the three vectors $\bb_\alpha = \tfrac{1}{2} \eps_{\alpha\beta\gamma} (\bk_\beta - \bk_\gamma)$ ($\eps_{\alpha\beta\gamma}$ is the fully antisymmetric tensor), also shown in Fig.~\ref{fig-lattice}(a). We note that any phase shifts that appear in general in the arguments of the three cosines in Eq.~(\ref{eq-lg1}) can be eliminated by a proper choice of the origin. 

We have studied the band structure of each of the two uncoupled lattices $V_{g/e}^{(1)} (\bs r)$ from first principles using the method and code published by Walters and coworkers \cite{walters13}. The Bloch states were computed and used to construct a localized basis spanned by the maximally-localized generalized Wannier functions~\cite{marzari2012}. Knowledge of the Wannier functions in turn enables one to compute the parameters of a faithful tight-binding model describing dynamics in the lowest bands for each lattice. The band structure calculation also signals the limits of validity of this tight-binding model, see also Ref.~\cite{lee09}. We will consider in the following the (arbitrary) criterion for the validity of this model: the width $W_s$ of the lowest $s$-band is one order of magnitude lower than the gap $\Delta_{sp}$ separating this band from the higher lying $p$-band.

\begin{figure*}
\centering
\includegraphics[width=0.72\textwidth]{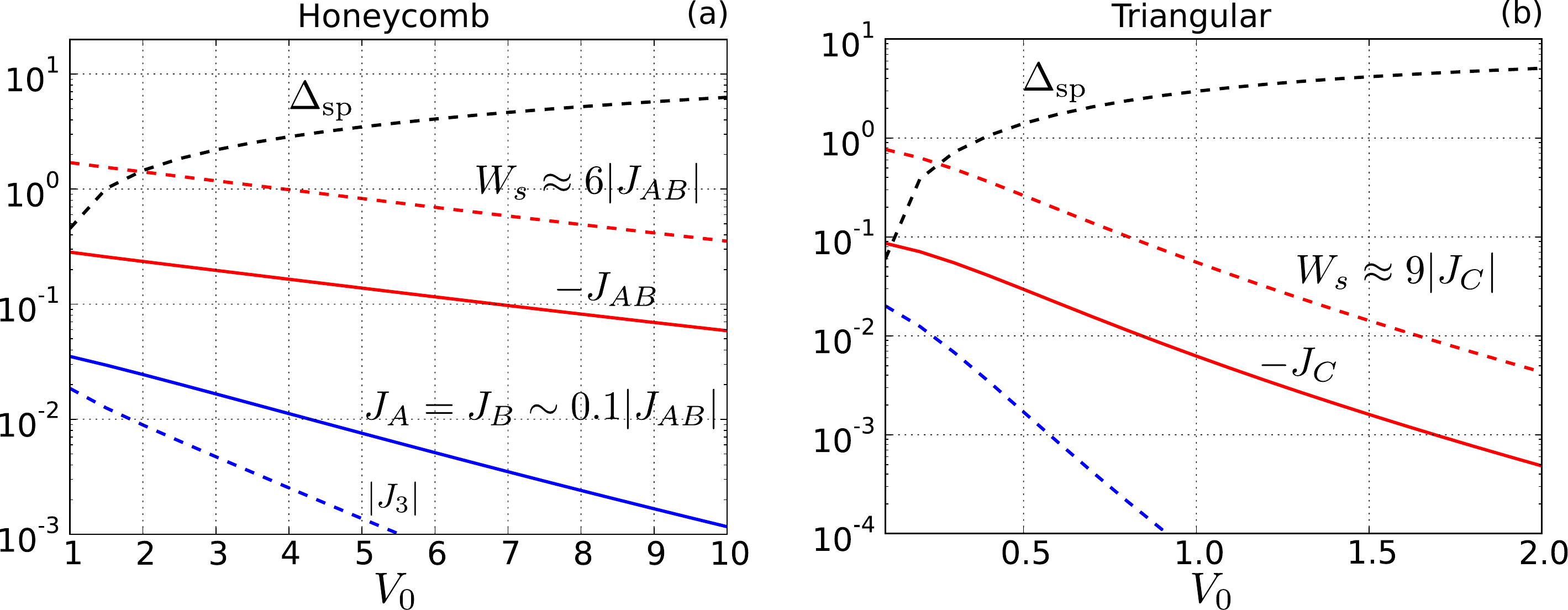}
\caption{\label{fig-hextri} (Color online) Band structure parameters for the main optical lattice shown in Fig.~\ref{fig-lattice}(c), as a function of the potential strength $V_0$: (a) The honeycomb lattice felt by the $e$ states, and (b) the triangular lattice felt by the $g$ states.  Both panels show the width of the lowest Bloch band $W_s$ ($W_s = 6|J_{AB}|$ for the honeycomb lattice and $W_s = 9|J_C|$ for the triangular lattice), the band gap $\Delta_{sp}$ separating it from the higher band and the hopping amplitudes. All quantities are energies, expressed in units of the recoil energy $E_R$.}
\end{figure*}

We start with the honeycomb lattice potential $V_e^{(1)} (\br)$ felt by atoms in state $e$, and present the results of the full band-structure calculations in Fig.~\ref{fig-hextri}(a). The tight-binding model relevant to the two lowest energy bands -- analogous to the well-known bands of graphene that touch at the Dirac points \cite{neto2009a} -- is parameterized by a NN hopping matrix element $J_{AB}$ connecting inequivalent sites [red line in Fig.~\ref{fig-hextri}(a)] and a NNN hopping matrix element connecting equivalent sites, $J_{A}=J_{B}$ [blue line in Fig.~\ref{fig-hextri}(a)]. The NNN hopping amplitude is positive and at least an order of magnitude smaller that the NN hopping. For comparison, the amplitude of the third order transition is also included: $|J_3|$ is the absolute value of the matrix element connecting a given site to the diametrically opposite site across a honeycomb cell. This element is negative, and is the largest of all neglected higher-order contributions. The lowest two bands have an energy width $W_s$ (given by $W_s = 6|J_{AB}|$ in the tight-binding approximation), which is plotted as a red dashed line to compare it to the bandgap $\Delta_{sp}$ separating the ground $s$ bands from the higher lying $p$ bands (black dashed line). We see that a two-band approximation is well justified for $V_0 \gtrsim 5~E_R$. This range also corresponds to $\vert J_3 \vert \lesssim J_A/10$. We conclude that for $V_0 > 5\,E_R$, the ground band is well isolated from the higher-lying ones and that a faithful two-site tight-binding model can be formulated including only NN and NNN transitions. 

The triangular lattice potential felt by the $g$ state is given by $V_g^{(1)} (\br)=-V_e^{(1)} (\br)$, whereby the minima and maxima exchange their positions. The corresponding C sites are separated by higher barriers than in the honeycomb lattice, and the distance between \emph{nearest-neighboring} sites of the triangular lattice is equal to the distance between \emph{next-nearest-neighboring} sites of the honeycomb lattice. As a consequence, for a given depth $V_0$ the tunneling rates in the triangular lattice are drastically smaller than in the honeycomb lattice. Figure \ref{fig-hextri}(b) summarizes the numerical results of the Wannier-structure study for the case of the triangular lattice. The red line shows the absolute value of the (negative) NN hopping element $J_C$, which is compared to the bandgap (black dashed line) and NNN hopping (blue dashed line). We see that for $V_0 > 0.5\,E_R$ the tight-binding model is well justified. Overall, the validity of single-band and tight-binding approximations are determined by the honeycomb lattice parameters. In the range $V_0>5~E_R$, tunneling between $C$ sites in the triangular lattice is weaker by orders of magnitude than for $A$ or $B$ sites. Although this seems like a serious concern for an experimental implementation, we will see later that introducing the auxiliary lattice cures this imbalance.

\subsection{Coupling the two sublattices}\label{section:coupling}

We now connect the two sublattices by a laser resonantly coupling the two internal states $g$ and $e$ and thereby induce hopping between the otherwise unconnected sublattices. We call this configuration the "hybrid lattice" in the following.  Following Ref.~\cite{Jaksch:2003}, we express the laser-assisted hopping matrix element between $A$ and $C$ sites, respectively hosting states $e$ and $g$, as
\begin{equation}
\label{eq:rabi}
  J (\br_A, \br_C) = \frac{\hbar\Omega}{2} \int \! d^2 r\, w_A (\br - \br_A)\, \e^{\i \bp\cdot\br}
  w_C (\br - \br_C),
\end{equation}
where the real-valued Wannier functions $w_A$ and $w_C$ are centered at their respective lattice sites $\br_A$ and $\br_C$. Here, $\Omega$ is the Rabi frequency characterizing the strength of the light-atom coupling, and $\bp$ is the recoil momentum transferred to the hopping atom. Since the product of the Wannier functions is well-localized near the midpoint of the line connecting the two sites, laser-induced hopping matrix elements are well approximated by
\begin{align}
&J (\br_A, \br_C) = J_{AC} \cdot\e^{\i \bp \cdot (\br_A + \br_C) / 2},\label{hopping_phase}
\end{align} 
where $J_{AC}$ is independent of $\bp$~\cite{Jaksch:2003}. By symmetry, one obtains $J (\br_A, \br_C) = J (\br_C, \br_A)^\ast$, and equivalent expressions for the hopping between $B$ and $C$ sites. 

Importantly, the hopping matrix elements in Eq.~\eqref{hopping_phase} contain space-dependent phases determined by the laser's wave vector $\bs p$. The sum of the phase factors along the boundary of a region $\triangle$ can be identified with the circulation of a synthetic vector potential penetrating the region $\triangle$. In the following, we use the term ``flux'' through a region $\triangle$ to refer to the synthetic flux given by the circulation of these phases along the boundary $\partial \triangle$. In the present work, we seek for a lattice configuration that gives rise to topological band structures with non-zero Chern numbers~\cite{Thouless1982,Kohmoto:1985}. As realized by Haldane~\cite{Haldane:1988}, a necessary condition to generate such topological band structures is to build a model that explicitly breaks time-reversal symmetry. Thus, a simple way to identify whether our hybrid honeycomb-triangular lattice indeed supports potentially non-zero Chern numbers is to examine its behavior under time reversal. 

\begin{figure}
\centering
\includegraphics[width=0.5\textwidth]{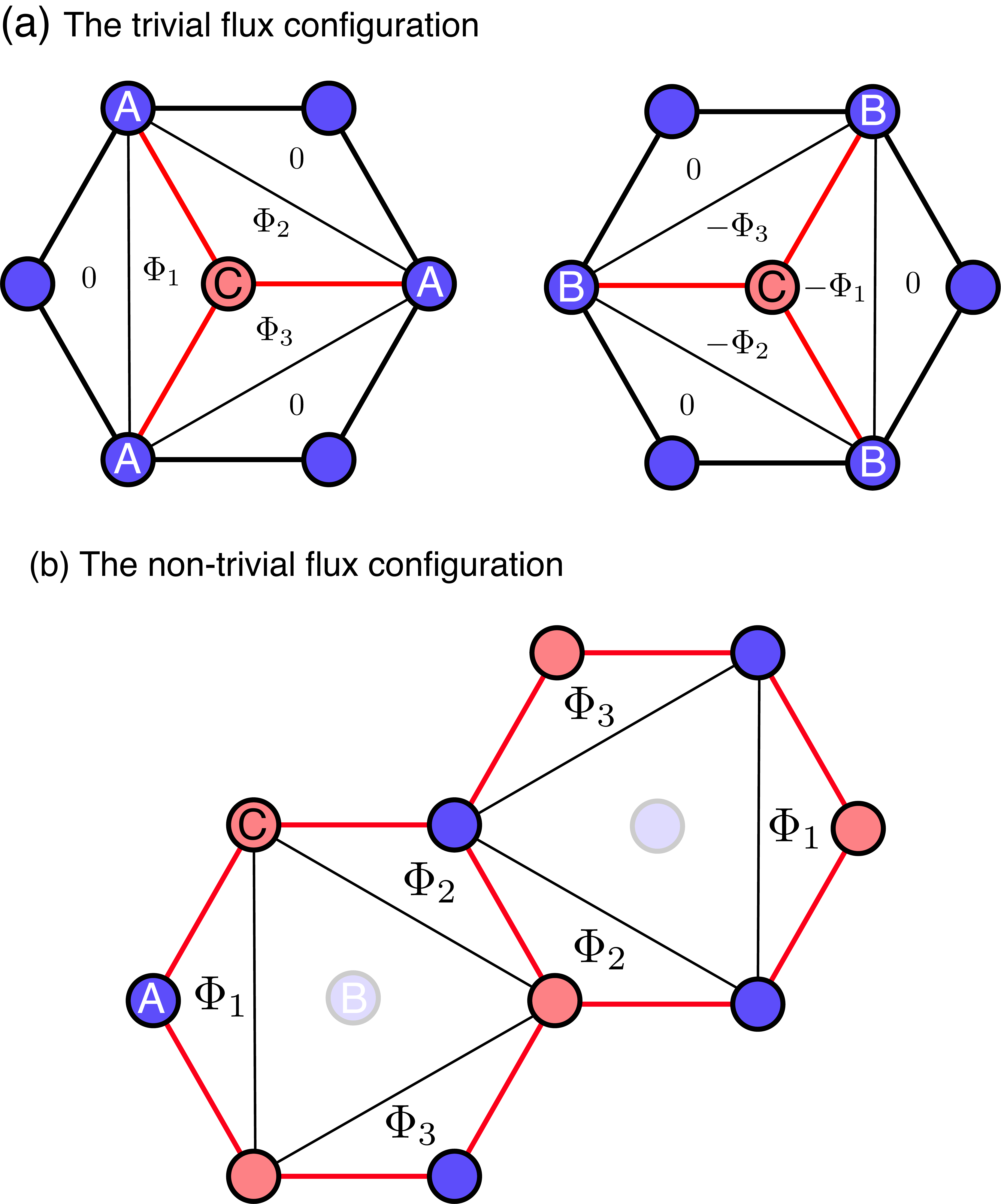}
\caption{ \label{fig-flux} (Color online) (a) Flux pattern for the hybrid honeycomb-triangular lattice with laser coupling. The phases accompanying the laser-assisted hopping \eqref{hopping_phase} lead to non-zero fluxes $\pm \Phi_{1,2,3}$ within the triangular subplaquettes shown on the figure. The fluxes are given by $ \Phi_1= \bs p \cdot \bs a_3/4\pi, \Phi_2= - \bs p \cdot \bs a_2/4\pi, \Phi_3= \bs p \cdot \bs a_1/4\pi$, where $\bs p$ is the recoil momentum and $\bs a _{1,2,3}$ are defined in Fig. \ref{fig-wannier}. Time reversal changes $\Phi_j \rightarrow - \Phi_j$, and therefore merely transforms A sites to B sites (and {\it vice-versa}). Since A and B are related by a discrete symmetry ($\Pi$ rotation around the axis perpendicular to the lattice plane), we conclude that the laser-coupled hybrid lattice does not break time-reversal symmetry. (b) Flux pattern for the main lattice perturbed by the auxiliary lattice introduced in Section \ref{auxiliary}. The B sites are eliminated from the lowest energy band by a strong on-site perturbation. The resulting low-energy tight-binding model is no longer invariant under time-reversal.}
\end{figure}

We will now demonstrate that the hybrid lattice is actually \emph{invariant} under this transformation. We show the flux patterns obtained from Eq.~\eqref{hopping_phase} in Fig.~\ref{fig-flux}(a) for two chosen subplaquettes patterns: the first one is spanned by A-C and A-A links, and the other by B-C and B-B links. Time reversal affects the lattice by reversing the sign of the fluxes. From the flux patterns shown in Fig.~\ref{fig-flux}(a), it is clear that this transformation leaves the honeycomb sublattice unchanged up to a discrete rotation. A similar analysis applies to other subplaquettes configurations, such as those spanned by C-C links. From this analysis, we conclude that the laser-coupled hybrid lattice remains time-reversal invariant even with laser-assisted tunneling, due to the high degree of symmetry between the A and B sites of the honeycomb sublattice. This also suggests that breaking this symmetry (e.g.\ by adding an onsite perturbation acting on the B sites only) will naturally generate a configuration that will change under time-reversal. This is the situation that we are going to analyze in the following Section.

\section{Adding the auxiliary lattice: Building the Haldane model}\label{auxiliary}

\subsection{The auxiliary lattice}

In order to remove the $A/B$ symmetry of the honeycomb lattice, we introduce an auxiliary lattice $V^{(2)}_{g/e}$ produced by two additional beams with wave vectors $\bk_1$ and $\bk_2$ and in-plane polarizations [see Fig.~\ref{fig-lattice}(b)]. The additional lasers are described by electric fields $ {\bs E}_{1}^{(2)} = E_2 \, \e^{\i\bk_1\cdot\br} \left( \tfrac{1}{2}, \frac{\sqrt{3}}{2} \right)$ and $ {\bs E}_{2}^{(2)} = E_2 \, \e^{\i\bk_2\cdot\br + \i \theta} \left( \tfrac{1}{2}, -\frac{\sqrt{3}}{2} \right)$, where $\theta$ is the relative phase shift between the two fields. Their coherent superposition produces a standing wave that adds incoherently to the existing main lattice $V^{(1)}_{g/e}$ due to the orthogonality of polarizations, {\it i.e.} $\vert {\bs E}_{\rm tot} (\br) \vert ^2 =\vert {\bs E}^{(1)} (\br) \vert ^2 +\vert {\bs E}^{(2)} (\br) \vert ^2 $. The potentials corresponding to this auxiliary lattice read
\begin{equation}
V^{(2)}_{g/e} (\br) ^2 =  \mp V_2  \left[2 - \cos \left(\bb_3 \cdot \br - \theta\right)\right].
\end{equation} 
The relative phase $\theta$ can not be eliminated by a change of origin, and -- together with the beam amplitudes --  allows one to move the position of the auxiliary lattice relative to the lattice $V^{(1)}_{g/e}$ and tune its depth. This way, the overall lattice geometry $V^{(1)}_{g/e} + V^{(2)}_{g/e}$ can be tuned. For a strong enough potential, the B sites of the primary honeycomb lattice $V^{(1)}_e$ are effectively eliminated from the dynamics [see Fig.~\ref{fig-lattice}(d)] leading to the desired laser-coupled honeycomb lattice illustrated in Fig.~\ref{fig-intro}(c).

\subsection{Perturbative analysis}

\begin{figure*}
\centering
\includegraphics[width=0.86\textwidth]{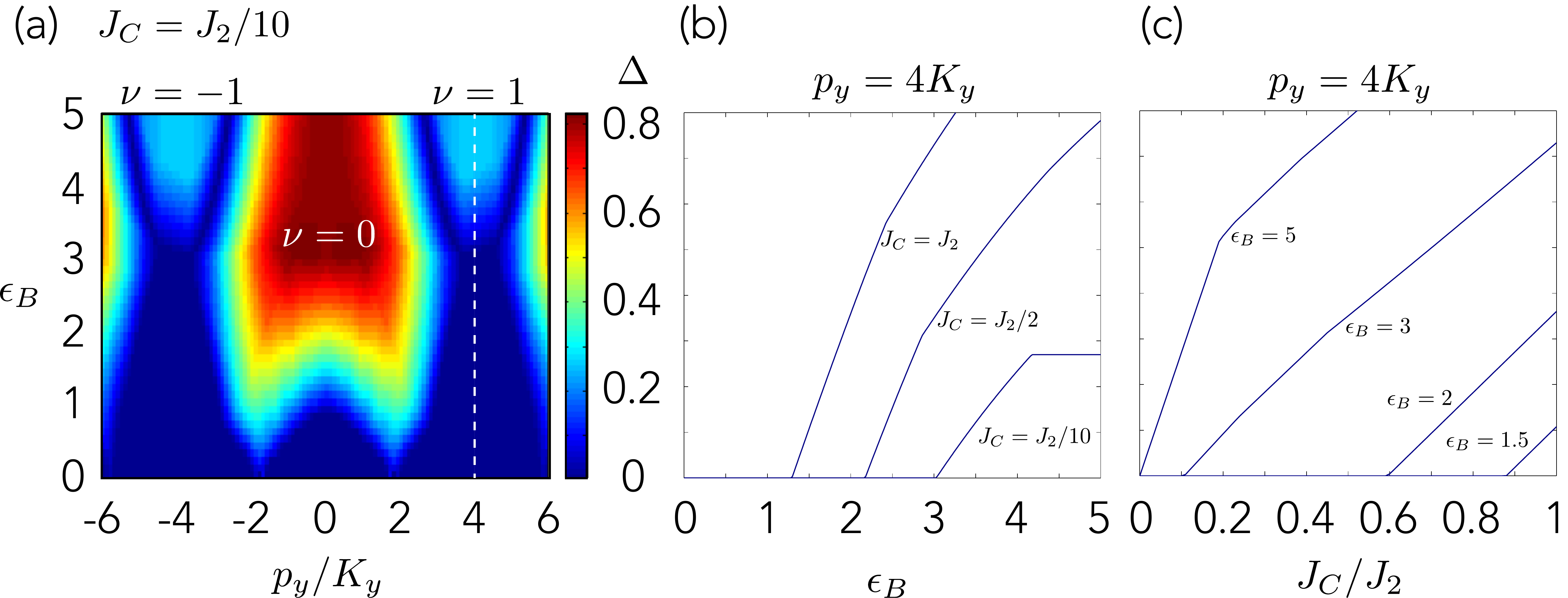}%
\caption{\label{fig-pert} (Color online) Band structure of the simplified tight-binding model \eqref{simple_ham} for the laser-coupled hybrid lattice. (a) Size of the main gap $\Delta$ as a function of the onsite perturbation $\epsilon_B$ and the recoil momentum $p_y$ for $J_2=0.3\,J$, $p_x=0$ and $J_C=J_2/10$. The large gap centered around $\bs p=0$ is associated with a zero Chern number, $\nu=0$, whereas the small gaps at $p_y \approx \pm 4 K_y$ are associated with the nontrivial Chern numbers $\nu=\pm 1$. Here, the gap $\Delta$ and the perturbation $\epsilon_B$ are expressed in units of the NN hopping $J$. (b) Cut through the diagram (a) for $p_y=4 K_y$ and increasing values of the ratio $J_C/J_2$. (c) Size of the main gap as a function of $J_C/J_2$ for increasing values of the perturbation $\epsilon_B$. Here $K_y=2 \pi / a 3\sqrt{3}$ and $a=2 \lam/ 3 \sqrt{3}$ is the lattice spacing of the primitive honeycomb lattice.}
\end{figure*}

To gain insight into the influence of the auxiliary lattice, let us first discuss the behavior of the band structure of the main hybrid lattice for a \emph{weak} onsite perturbation. We simplify the analysis by assuming that the system is well represented by a tight-binding model for the ground bands and by reducing the number of parameters. We take the absolute value of the (negative) NN hopping amplitude as the unit of energy so that $J_{AB} = -1$, and set the laser-induced hopping $J_{AC} = J_{BC} = 1$. The intra-sublattice NNN hoppings are considered to be uniform over the lattice $J_2 = J_A = J_B$, and $J_C \ll J_2$. We then add a perturbation $\epsilon_B$ that modifies the onsite energy of all the B sites, modeling the effect of a \emph{weak} external potential that aims to lift the spectrum degeneracy and open gaps, but still remains weak enough not to perturb significantly the band structure of the uncoupled lattices. 

The tight-binding model is then defined by the momentum space Hamiltonian
\be
H(\bs k )= \begin{pmatrix} 
J_2 f(\bs k + \tfrac{1}{2}\bs p) & g (\bs k) & h(\bs k) \\ 
g^* (\bs k) & J_2 f(\bs k + \tfrac{1}{2}\bs p) + \epsilon_B & h^* (\bs k) \\ 
h^* (\bs k) & h (\bs k) & J_Cf(\bs k-\tfrac{1}{2}\bs p)
\end{pmatrix}\label{simple_ham}
\ee
where
\begin{align}
&f(\bs k)= 2 \sum_{j=1}^{3} \cos \left ( \bs a_j \cdot \bs k \right ), \notag \\
&g(\bs k)= - \sum_{j=1}^{3} e^{ \i \bs \delta_j \cdot (\bs k + \bs p/2)}, 
\quad h(\bs k)= \sum_{j=1}^{3} e^{ -\i \bs \delta_j \cdot \bs k }, \notag 
\end{align}
where the vectors $\boldsymbol{a}_{j}$ and $\boldsymbol{\delta}_{j}$ are defined in the caption of Fig.~6.
We have analyzed the band structure through a direct diagonalization, varying the parameters in a wide range. In general one finds three bands, whose topological character can be established by computing the Chern number through the numerical method of Ref.~\cite{fukui2005}. For $ \epsilon_B = 0$, the two lowest bands touch at the Dirac points for any value of ${\bs p}$. A finite $ \epsilon_B > 0$ opens a gap $\Delta$ separating these two bands. 

Figure \ref{fig-pert}(a) shows the magnitude of $\Delta$ for $J_C \ll J_2$, and indicates the opening of gaps of different nature as the perturbation $\epsilon_B$ is increased. The figure also indicates the Chern number $\nu$ associated with the lowest isolated band. The Chern number has been computed using the method of Ref. \cite{fukui2005}, which is based on an efficient discretization of the Berry's curvature inspired by lattice gauge theory. A large trivial gap ($\nu=0$) is first opened for small $\epsilon_B$ around the time-reversal-invariant configuration ($\bs p=0$). For large $\epsilon_B$, nontrivial gaps with Chern numbers $\nu = \pm 1$ open at finite $\bs p \ne 0$. Non-zero Chern numbers $\nu = \pm 1$ imply that the lowest energy band is associated with a non-trivial topological order \cite{Goldman:2013njp}: setting the Fermi energy  within the gap leads to a Chern insulating phase, characterized by chiral edge modes \cite{GoldmanDalibard:2012,GoldmanEdge:2013,Stanescu2010,Liu:2010,Scarola:2007}. We identify these nontrivial Chern insulating phases with those that were previously reported in Ref.~\cite{Goldman:2013njp}, namely, the phases resulting from the Haldane-like model obtained by only considering the presence of A and C sites (i.e. removing the B sites of the hybrid lattice). The opening of this topological gap is further analyzed in Figs.~\ref{fig-pert}(b)-(c), by varying the hopping $J_C$. These plots show that even for unrealistically large hopping between the C sites of the primitive triangular lattice ($J_C \sim J_2$), a very large onsite perturbation $\epsilon_B$ is required to generate a topological phase. 

\begin{figure*}
\centering
\includegraphics[width=0.86\textwidth]{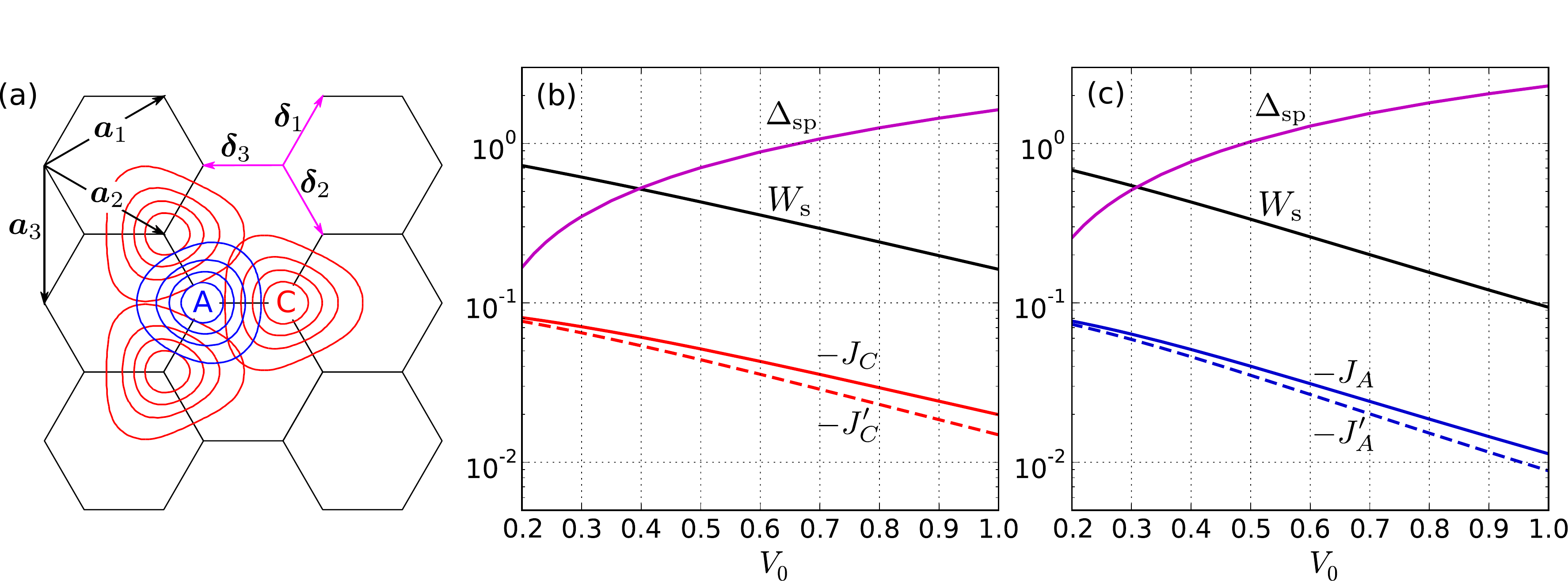}%
\caption{\label{fig-wannier} (Color online) (a) A fragment of the honeycomb lattice AC, corresponding to the total potential combining the main and auxiliary lattices, simultaneously showing the elementary translation vectors and the shape of the Wannier orbitals. The contour levels are drawn at $0.2, 0.4, 0.6$, 
and $0.8$ times each orbital's maximum value. The vectors are given by $\bs \delta_1= a/2 (1, \sqrt{3})$, $\bs \delta_2= a/2 (1, -\sqrt{3})$, $\bs \delta_3= a(-1,0)$, $\bs a_1 = \bs \delta_1-\bs \delta_3$, $\bs a_2 = \bs \delta_2-\bs \delta_3$ and $\bs a_3 = \bs a_2- \bs a_1= \bs \delta_2-\bs \delta_1$, where $a=2 \lam / 3 \sqrt{3}$ is the primitive lattice spacing. Panels (b) and (c) show the 
width of the lowest Bloch band $W_s$, the band gap separating from the higher band $\Delta_{sp}$ 
and the Hubbard parameters for the respective sublattices C and A. All quantities in (b)-(c) are energies, expressed in units of the recoil energy $E_R$.}
\end{figure*}

\subsection{Band structure calculations and tunneling parameters}

For very large $\epsilon_B \gg J_{AB}$, the perturbative analysis presented above breaks down as the lattice geometry becomes strongly distorted. We have performed a full band-structure calculation based on the full potential to re-evaluate the proper parameters for the tight-binding model of the hybrid lattice in the lowest band. Although the auxiliary lattice leads to onsite energies that are the same for all equivalent sites (A, B or C), it does not respect the original triangular point symmetry of the primary potential landscape and affects the potential landscapes away from the maxima or minima. As a consequence, for arbitrary $\theta$, the hopping amplitudes between neighboring potential minima are generally direction dependent. Numerical work reveals that choosing the values $\theta = \pi/6$ and $V_2=3  V_1$, as shown in Fig.~\ref{fig-lattice}(d), is optimal to preserve -- albeit approximately -- the triangular point symmetry of the potential landscape. The calculations presented in the following are performed using these values.

Figure \ref{fig-wannier} summarizes the results. Panel (a) shows a fragment of the lattice. The blue (resp.\ red) contour lines depict the shapes of the calculated real-valued maximally-localized Wannier functions on a single site A (resp.\ three surrounding sites C). We see that the Wannier orbitals have rounded triangular shape that follows the shape of the potential well in the vicinity of the potential minima. Although not immediately conspicuous in the contour plots, the Wannier orbitals \emph{do not} have the full $D_3$ symmetry of the equilateral triangle; instead, they are only symmetric with respect to reflection in the $x$ axis. This is a consequence of the striped auxiliary interference pattern and is reflected in a slight directional dependence of hopping amplitudes, see Fig.~\ref{fig-intro}(c).

The calculated hopping matrix elements and characteristics of the energy bands are shown in panels (b) and (c) of Fig.~\ref{fig-wannier}. All the plotted parameters have the dimensions of the energy and are expressed in terms of the recoil energy $E_R$. The two panels correspond to the different triangular sublattices, and are completely analogous. Thus, we restrict the discussion to the behavior of $g$  atoms shown in panel (b). The full and dashed red lines show the dependence of the hopping amplitudes between NNN sites of type C. As expected, these hopping elements display a weak directional dependence. Thus, transitions connecting two sites in the $\pm \ba_3$ direction ($J_C'$) are slightly weaker than transitions connecting neighboring C sites in the $\pm \ba_{1(2)}$ directions ($J_C$). The full black and purple lines indicate, respectively, the dependence of the width of the lowest $s$ band $\Delta_s$ and the band gap $\Delta_{sp}$ to the higher $p$ band. We have also verified that higher order hopping transitions are negligible. Using the same criterion as before ($\Delta_{sp} \gtrsim 10~W_s$), we conclude that a single-band tight-binding approximation becomes justified as soon as the modulation strength exceeds $V_0 \approx 1\,E_R$. 

We stress that the obtained tunneling parameters are now all similar in magnitude, unlike the situation without auxiliary lattice, and that they only weakly depend on the direction despite the absence of triangular point symmetry in the strict sense. For example, the choice $V_0 = 1\, E_R$ leads to values
\begin{equation}
\begin{split}
\label{eq:vals}
  J_A &= - 0.011\,E_R, \quad J_A' = -0.009\,E_R,\\
  J_C &= -0.020\,E_R, \quad J_C' = -0.015\,E_R.
\end{split}
\end{equation}
We also verified that the same conclusion applies to the inter-sublattice NN transitions, that is, the hopping amplitudes show only a weak dependence on the direction of the AC link given by $\bm{\delta}_{1,2,3}$.

\subsection{Tight-binding model}

In the tight-binding approximation, the model is represented by the $\bk$-space Hamiltonian 
\begin{equation}
\label{eq-matrix}
\mathcal{H}(\bk)=\left(\begin{array}{cc}
  F(J_A, J_A', \bk + \tfrac{1}{2}\bp) & 
  J_{AC}\, h(\bk)\\
  J_{AC}\, h^\ast(\bk) & 
  F(J_C, J_C', \bk - \tfrac{1}{2}\bp)
\end{array}\right),
\end{equation}
where
\begin{align*}
  F(J, J', \bk) &= 2J \sum_{j=1}^2 \cos(\bk\cdot\ba_j) + 2J' \cos(\bk \cdot \ba_3),
\end{align*}
and the recoil momentum $\bp$ enters the arguments of these functions as a shift in the reciprocal space.

\begin{figure*}
\centering
\includegraphics[width=0.64\textwidth]{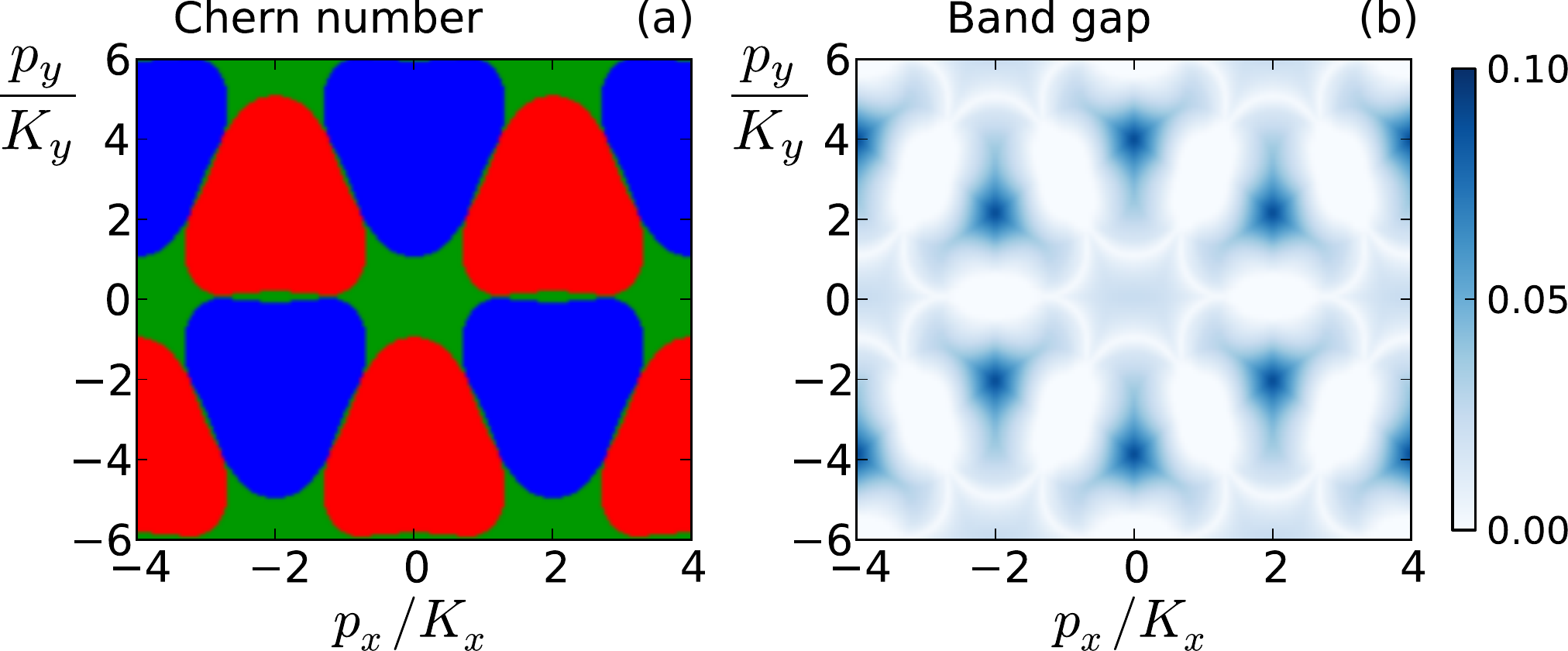}
\caption{\label{fig-tba} (Color online) Topological band structure in the tight-binding regime. The left panel shows the dependence of the lower-band Chern number on the Cartesian components of the recoil momentum. The latter are expressed in terms of the vector $\bm{K} = (2\,\bb_1 + \bb_2) / 3 = (\sqrt{3}\pi/\lam, \pi/\lam)$ pointing to a Dirac point of the reciprocal lattice. Red/blue areas denote Chern numbers $\pm 1$, green areas are topologically trivial. The right panel shows the gap separating the two bands with opposite Chern numbers. The band gap is expressed in units of the recoil energy $E_R$.}
\end{figure*}

We calculate the band and topological structure numerically using the realistic parameter values obtained from the band structure modelling at the potential modulation strength $V_0 = 1\, E_R$. NNN hopping amplitudes are listed in Eq. \eqref{eq:vals} and take values in the vicinity of $J_A,J_C \approx -0.015\,E_R$. Guided by our previous work \cite{Goldman:2013njp}, we set the strength of the laser-assisted NN transitions to $J_{AC} \approx 3|J_{A}| = 0.050\,E_R$, which corresponds to using the Rabi frequency $\hbar\Omega \approx E_R$ in Eq.~\eqref{eq:rabi}. Figure~\ref{fig-tba}  shows the Chern number of the lowest band, which has been numerically computed using the method of Ref. \cite{fukui2005}. This confirms that the topological phases are indeed readily accessible in this regime. The Chern number patterns are periodic in $\bp$ with a hexagonal unit cell twice the size of the ordinary Brillouin zone. This is the consequence of the fractional argument $\bp / 2$ entering the matrix elements of the Hamiltonian matrix (\ref{eq-matrix}). 

The obtained phase diagram in Fig.~\ref{fig-tba}(a) is dominated by areas corresponding to topologically nontrivial regimes. A possible experimental detection of topological phases requires that the two bands characterized by Chern numbers $\pm 1$ are separated by substantial band gaps. Panel (b) shows that the band gap can exceed the coupling strength $J_{AC}$ or, in the best cases, even approach $\Delta=2\, J_{AC} = 0.10\,E_R$.  We also note that the gap attains the maximum values on a lattice spanned by the vectors $2\bm{K} = (2K_x, 2K_y)$ and $2\bm{K}' = (2K_x, -2K_y)$ in the $\bp$ plane. Thus, the six maximum-bandgap points closest to the origin correspond to the recoil momenta $\bp = 4\pi/\lam$ and are nearly reachable employing the largest possible recoil momenta $\bp_{\text{max}} = 2\pi / \lambda_{ge}$ with the resonant wavelength $\lambda_{ge} = 578\,\mathrm{nm}$.   

\subsection{Detection of the Chern insulating phase}

Different methods to detect topological order in cold-atom setups have been recently proposed. Two routes are generally envisaged: (a) measure the Chern number~\cite{Alba:2011,Price:2012,Abanin:2012,Dauphin:2013,Liu:2013} or (b) detect the presence of chiral edge modes ~\cite{GoldmanDalibard:2012,GoldmanEdge:2013,Stanescu2010,Liu:2010,Scarola:2007}.

In two-band models, described by the general Hamiltonian
\be
H(\bs k)= \epsilon (\bs k) \hat{1}_{2\times 2}+ \bs{d} (\bs k) \cdot \bs{\hat \sigma},
\ee
such as the Haldane-like system considered here, the Chern number is directly related to the vector field $\bs{d} (\bs k)$, through the winding-number expression
\be
\nu=\frac{1}{4 \pi} \int_{\mathbb{T}^2} \frac{\bs{d}}{\vert \bs d \vert^3} \,   \cdot \biggl ( \partial_{k_x} \bs{d} \times  \partial_{k_y} \bs{d}    \biggr) \txt{d}^2 \bs{k},
\ee
which counts the number of times the vector $\bs{d} (\bs k)/\vert \bs d \vert$ covers the unit sphere as $\bs k$ is varied over the Brillouin zone. Following Ref. \cite{Alba:2011}, the vector $\bs{d} (\bs k)$ could be reconstructed through spin-resolved time-of-flight measurements, allowing for a ``pixelated" measure of the Chern number. 

More generally, it has been shown that releasing a Fermi gas initially prepared in a Chern insulating phase and acting on the cloud with an external linear potential (i.e. a synthetic ``electric" field $\bs E$) leads to a clear transverse (Hall) drift of the cloud: measuring the center-of-mass displacement in the direction transverse to the field $\bs E$ provides a direct measure of the Chern number $\nu$ \cite{Dauphin:2013}. Alternatively, signatures of the Berry's curvature $\mathcal{F}(\bs k)$ could be detected through Bloch oscillations \cite{Price:2012,Abanin:2012}, offering an alternative way to reconstruct the Chern number $\nu \approx (1/2 \pi i) \sum_{\bs k} \mathcal{F} (\bs k)$. 

Edge modes could be directly visualized through the methods of Ref. \cite{GoldmanDalibard:2012}, which allows to detect the propagation of edge states on a dark background (i.e. in a region unoccupied by the many bulk states). Alternatively, the linear dispersion proper to chiral modes could be identified through spectroscopy measurements~\cite{GoldmanEdge:2013,Stanescu2010,Liu:2010}.

\section{Conclusions}\label{conclusion}

In summary, we have introduced and analyzed a realistic scheme to realize a Chern insulator using cold atoms. In this scheme, one exploits: (i) the presence of a long-lived excited state in addition to the actual ground state, which is characteristic to alkali-earth or Ytterbium atoms, and (ii) the existence of a frequency range where the polarizabilities of the two relevant states differ in sign. This allows to exploit both intensity maxima and minima of an optical lattice to trap the two internal states, simultaneously avoiding heating from spontaneous emission. Based on first-principle calculations, we validate the applicability of the tight-binding approach in certain parameter regimes, and demonstrate the emergence of a generalized Haldane model, with laser-induced complex nearest neighbor transitions and natural real-valued next-nearest neighbor transitions. We show that topological phases are indeed readily accessible, with the topological bandgaps on the order of $0.1\,E_R \sim 100\,\text{Hz}$, indicating that the topological properties could be detected at sufficiently low temperatures $\sim \text{nK}$ using currently existing proposals based on Chern-number measurement~\cite{Alba:2011,Price:2012,Abanin:2012,Dauphin:2013} or edge-state detection~\cite{GoldmanDalibard:2012,GoldmanEdge:2013,Stanescu2010,Liu:2010,Scarola:2007}. Finally, we emphasize that our proposal to implement the Haldane model using long-lived excited states follows an earlier proposal \cite{Jaksch:2003,Gerbier:2010} to realize the paradigmatic Hofstadter model~\cite{Hofstadter:1976}, suggesting that the versatility of this scheme could be further exploited to realize other lattice systems of interest.


\begin{acknowledgments}
N.G.\ is supported by the Universit\'e Libre de Bruxelles (ULB). This research was also funded by the European Social Fund under the Global Grant measure, by Ville de Paris under the Emergences program [AtomHall] and by the European Research Council under the EU Seventh Framework Program (FP/2007-2013) [StG MANYBO GA 258521]. Discussions with J.~Dalibard, J.~Beugnon, A.~Eckardt, G.~Juzeli\={u}nas and J.~Ruseckas are gratefully acknowledged. E.A.\ thanks T.~H.~Johnson for correspondence and providing a pre-publication version of the Wannier code.
\end{acknowledgments}


\bibliographystyle{apsrev}

\end{document}